# Nanoindentation-induced evolution of atomic-level properties in silicate glass: Insights from molecular dynamics simulations

Linfeng Ding[a+]*, Ranran Lu[a+], Lianjun Wang[a], Qiuju Zheng[b], John C. Mauro[c] and Zhen Zhang[d]*

[a]*State Key Laboratory for Modification of Chemical Fibers and Polymer Materials, Engineering Research Center of Advanced Glass Manufacturing Technology, Ministry of Education, Donghua University, Shanghai 201620, China*

[b]*School of Materials Science and Engineering, Qilu University of Technology, Jinan 250353, China*

[c]*Department of Materials Science and Engineering, The Pennsylvania State University, University Park, PA 16802, USA*

[d]*State Key Laboratory for Mechanical Behavior of Materials, Xi'an Jiaotong University, Xi'an 710049, China*

[+]These authors contributed equally to this work
*Corresponding author: Linfeng.Ding@dhu.edu.cn; zhen.zhang@xjtu.edu.cn

**Abstract**

Indentation has been widely used for investigating the mechanical behavior of glasses. However, how the various microscopic properties (such as atomic structure and mechanics) of glass evolve from the immediate contact with the indenter to the far-field regions, and how these observables are correlated to each other remain largely unknown. Here, using large-scale molecular dynamics simulations, we investigate the response of a prototypical sodium silicate glass under shape contact load up to an indentation depth of 25 nm. Both the short- and intermediate-range structures are found to exhibit notable changes below the indent, indicating that indentation deformation induces a more disordered and heterogeneous network structure. In addition, we find that the indentation-induced changes of local properties all exhibit an exponential decaying behavior with increasing distance from the indent. Comparison of the characteristic decay lengths of these local properties indicates that the structural origins of shear flow and densification are the changes of the network modifier's coordination environment and the inter-tetrahedral connection, respectively. The decay of densification is considerably slower than that of shear strain, implying that the former might contribute more to the deformation at the far-field regions. Our findings not only contribute to an atomistic understanding of the indentation response of silicate glasses but also pave the way towards rational design of damage-resistant glassy materials.

# 1. Introduction

Oxide glasses, such as those used for the displays of electronic devices and pharmaceutical packaging, are often exposed to high stress via sharp contact during handling by the glassmaker and daily usage by the customer. This high stress will lead to the formation of surface flaws, e.g., cracks or indents, which in turn, act as stress concentrators, resulting in a significantly reduced practical strength of the materials.[1] Understanding the response of glasses under sharp contact loads is thus critical for the rational design of damage-tolerant glassy materials. Experimentally, indentation has been widely applied to study the mechanical response of glass undergoing high stress since this technique is not only highly controllable for introducing permanent deformation but also mimics real-life damage incidents in glasses.[2, 3, 4, 5, 6, 7] In particular, nanoindentation has become a popular technique in recent years for measuring the deformation of glass under sharp contact load with high precision.[8, 9, 10, 11, 12, 13]

During indentation of glass, both elastic and inelastic deformation occur, and the latter is responsible for the indentation sink-in (below the original contact surface) and pile-up[2] (above the original contact surface). Although not yet conclusive, previous experiments[5, 14, 15, 16, 17, 18, 19, 20, 21, 22, 23] and computer simulations[11, 24, 25, 26, 27, 28] have suggested that shear flow and densification are the two main mechanisms responsible for the inelastic deformation of glass under indentation; the respective contribution of each mechanism may depends on the shape/sharpness of the indenter.[5] A recent experimental study[13] on a series of silicate glasses found that the region above the original contact surface can be further separated into pile-up (close to the indent) and lift-up (far away from the indent). The origin of the latter part was attributed to the increasing width of the indenter as it is pushed into the glass (lateral-pushing force) which contributes to ~25% of the total volume above the original surface in a soda-lime silicate glass. However, the atomic-scale mechanism of the deformation at the far field from the indenter which leads to the formation of the lift-up region is still unknown.

To complement experimental investigations, computer simulations, particularly molecular dynamics (MD) simulations, have also been utilized for obtaining microscopic insights into the response of oxide glasses under nanoindentation. Both shear flow and densification have been observed in silica[29], borosilicate[11, 28, 30] and aluminosilicate[27] glasses. These nanoindentation simulations revealed that the primary mechanism responsible for densification is the decrease of network bond angles and that the amount of shear flow is governed by the content of modifiers (which reduce the network connectivity by disrupting Si-O-Si linkages). Structural changes on short ranges (e.g., the formation of oxygen triclusters and fivefold coordinated silicon[31]) as well as the change of medium-range structure (e.g., the connection of [SiO$_4$] tetrahedra into ring-like structures)[28, 30, 32], are also suggested to be important for understanding the deformation behavior of glass under nanoindentation.

Despite considerable efforts have been made to study the response of glass under the load of a sharp indenter, some fundamental questions are still not fully addressed. In this work, we highlight the question as to how the various atomic-level properties (structure, stress, strain, etc.) of the glass evolve from the immediate contact with the indenter to the far field (>30 nm) of the sample, and how these changes are coupled with the behavior of the network modifiers in the case of muti-component silicate glasses. To shed light on these questions, we use large-scale MD simulations to mimic the situation in which the glass is subjected to high stress as induced by a sharp indenter. Using a large-scale (~ 3.1 million atoms) simulation setup, we monitor and quantitatively characterize how the various local properties change as the indenter is pushed inside the glass. The findings of this work could help to establish the relationship between local composition, structure, and mechanical properties, hence allowing for a deeper microscopic understanding of the deformation mechanisms of the glass under a sharp contact load.

1.1 **Simulation details**

We performed MD simulations on a sodium silicate glass 25Na$_2$O-75SiO$_2$ (mole%, hereafter denoted as NS3). This composition was selected as a representation of multi-component silicate glasses for which the network modifying species such as Na play a decisive role in controlling the structure and properties of the glasses. The interactions between atoms were described by a two-body effective potential (named SHIK) recently proposed by Sundararaman *et al*.[33] This potential was parameterized using liquid structure from *ab initio* simulations and experimental data of glass density and elastic constants. Recent studies have shown that this potential gives a reliable description of the structural, mechanical, as well as surface properties of sodium silicate glasses.[34, 35, 36, 37, 38] Thus it can be expected that this potential also permits us to obtain useful insights on the deformation behavior of the NS3 glass under nanoindentation.

We first produced the NS3 glass sample via a melt-quench procedure. The initial sample consisting of half a million atoms was first melted at 3000K for 320 ps, a time span that is sufficiently long to ensure loosing memory of the initial atomic configuration and the equilibration of the melt. After that, the melt was cooled down to 300 K using a cooling rate of 1 K/ps. The glass sample was further relaxed at 300 K for 160 ps. The final glass sample has the dimensions of 48.4 nm, 36.3 nm, and 4.0 nm (in the *x*, *y*, and *z* directions, respectively). Throughout the melt-quench process, the simulations were carried out in the isothermal-isobaric (NPT) ensemble under zero pressure with periodic boundary conditions applied in all directions.

All the MD simulations were performed using the LAMMPS (Large-scale Atomic/Molecular Massively Parallel Simulator) package[39] at a fixed time step of 1.6 fs. Visualization of the atomic configurations was realized using the OVITO software[40].

Using the prepared glass sample, we first carried out uniaxial tensile simulations to calculate the elastic constants (Young's modulus, shear modulus, bulk modulus, and Poisson's ratio) of the glass. Specifically, the sample was elongated at a constant strain rate of 0.5 ns$^{-1}$ along the *x* direction, while zero stress was kept in the other two directions perpendicular to the loading axis. The stress-strain ($\sigma$-$\varepsilon$) curve was linearly fitted in the strain range of 0-0.001 to calculate the Young's modulus (*E*). The Poisson's ratio ($\upsilon$) was calculated by linearly fitting the transverse strain versus axial strain curve in the range of 0-0.01. The shear modulus (*G*) and the bulk modulus (*K*) were calculated by the equations[41]:

$$G = \frac{E}{2(1+\upsilon)} \text{ and } K = \frac{E}{3(1-2\upsilon)}.$$

Next, we performed nanoindentation simulations to investigate the response of the glass sample under a sharp contact load. In order to reduce the sample size effect on the nanoindentation test, we replicated the initial glass sample to approximately 145.0×74.4×4.0 nm³ (consisting of around 3.1 million atoms) at 600 K (the glass transition temperature as estimated from the inflection point of the enthalpy-temperature curve is around 1600 K). Then, the sample was quenched to 300 K with a cooling rate of 1 K/ps and subsequently maintained at this temperature for another 160 ps. This cooling-relaxation process permits minimization of the boundary effect introduced by replication of the initial glass sample. Similar strategies were used in previous simulation studies and the obtained results regarding the deformation behavior of glasses were found to be reasonable, see e.g., Refs.[27, 42].

Nanoindentation of the glass was simulated based on a setup originally proposed in Ref.[27], which has proven to be useful for understanding the indentation response of oxide glasses[27, 43]. A schematic representation of the simulation setup is shown in **Fig. 1**. The geometry of the indenter was controlled by the indenter angle $\theta$ and the indenter tip radius $R$. In line with a recent experimental study that used a cube-corner indenter to probe the indentation response of silicate glasses[13], we have chosen the indenter angle $\theta$ = 35.3°, and the tip radius $R$ was set to 0.1 nm. The periodic boundary condition along the $y$ direction was released to create a free surface on the top while the bottom layer with a thickness of 1 nm was fixed. The interaction between atoms and the indenter was defined by a spring constant of 32 GPa. During the loading and unloading processes, the indenter moved vertically along the $y$ direction at a constant speed of 50 m/s. At the maximum indentation depth of 25 nm, the indenter was held for 125 ps which is sufficiently long to ensure force convergence. We performed 6 parallel indentation tests at 6 different positions (equally spaced and made use of both sides along the $y$ direction of the sample). Error bars were evaluated as the standard error of the 6 parallel indentation tests. It is worth emphasizing that the system size considered in this study is large enough to ensure that the sample-to-sample fluctuations are negligible.

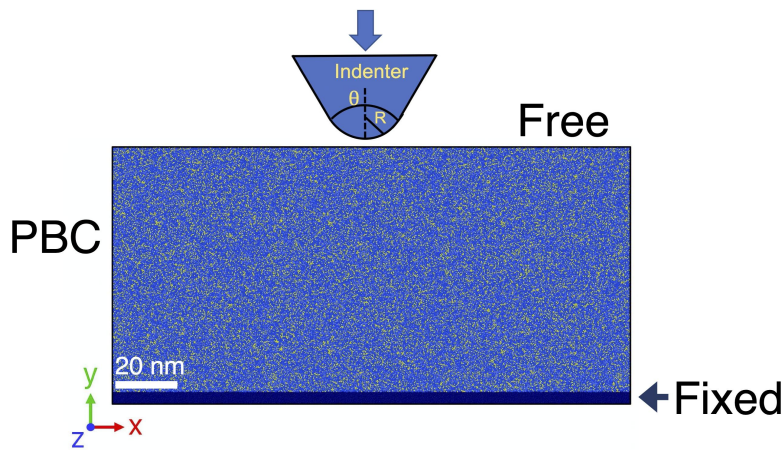

**Figure 1**. Schematic illustration of the nanoindentation setup.

2. Results and discussion

2.1 Properties of the bulk glass

**Table 1** presents the simulated elastic constants together with the experimental data[44] and the estimation from previous simulation studies using other interaction potentials[45, 46]. Firstly, one observes that the SHIK potential predicts the various elastic constants in good agreement with the experimental values, indicating that this potential is reliable for investigating the mechanical properties of the glass, in agreement with the finding of Zhang *et al*.[34]. Other popular interaction potentials, namely the Teter potential[46] and the Pedone potential[45], give a similar level of agreement in predicting the elastic constants relative to the SHIK potential.

It is worth mentioning that in addition to the high accuracy, another important reason for choosing the SHIK potential in this study is due to its high computational efficiency which is particularly advantageous when simulating large systems, say containing millions of atoms. These large systems are required to eliminate the finite size effect when studying the mechanical behavior of glasses such as fracture[34, 36] and indentation deformation[27].

Table 1. Properties of the NS3 glass compared with experimental data and MD simulations using different interatomic potentials.

|  | Young's modulus (GPa) | Bulk Modulus (GPa) | Shear Modulus (GPa) | Poisson's ratio | Density (g/cm$^3$) |
|---|---|---|---|---|---|
| Experiments[44] | 60 | 37 | 24 | 0.23 | 2.43 |
| Teter[46] | 55 | 37.8 | 22 | 0.26 | 2.56 |
| Pedone[45] | 58.1 | 38.5 | 23.3 | 0.23 | - |
| This work | 56.3 | 36.1 | 22.7 | 0.24 | 2.46 |

## 2.2 Macroscopic indentation response of the glass

Having demonstrated that the simulated glass sample is realistic, we move on to discuss the indentation response of the glass. Firstly, we present in **Fig. 2**(a) the force-displacement curve of the indenter during the indentation loading, holding, and unloading processes. In the loading stage, the indentation depth increases continuously, and the glass undergoes both elastic and inelastic deformation due to the applied sharp contact load. The glass sample exhibits stress relaxation during the holding stage as evidenced by the stress drop at the maximum loading depth. This notable relaxation behavior could be attributed to the high strain rate applied[27]. Upon unloading, the indented volume is partially recovered due to the reversible elastic deformation[11].

Next, we show in **Fig. 2**(b) the inelastic deformation profile along the *x* direction after unloading. This profile was determined by searching the atoms with the highest *y* coordinate every 5 Å along the *x* direction. One observes that the pile-up height is ~1.5 nm, which is around 6% of the maximum nanoindentation depth of 25 nm. This ratio agrees with the finding of a recent nanoindentation experiment on a soda-lime silicate glass ($72SiO_2$-$13Na_2O$-$10CaO$, mole%) which found an averaged pile-up height of 27.6 nm, i.e., ~5.5% with respect to the maximum penetration depth of 500 nm[13]. This good agreement indicates that the simulation setup applied in this study is reliable and the microscopic insights from these simulations should be of experimental relevance and usefulness for understanding the deformation behavior of the glass under a sharp contact load. We also note that we are not able to observe a clear signature of the far field lift-up region as suggested by a recent experimental work[13]. This discrepancy might be attributed to the much smaller system size and indentation depth considered in this work in comparison with the experimental counterparts, which make that the volume change in the far-field regions is insignificant. A recent simulation study of a model metallic glass[57] has shown that as the system size is increased to about 340 million atoms, an inelastic deformation mode (near the sample surface) similar to the lift-up observed in Ref.[13] can be recognized.

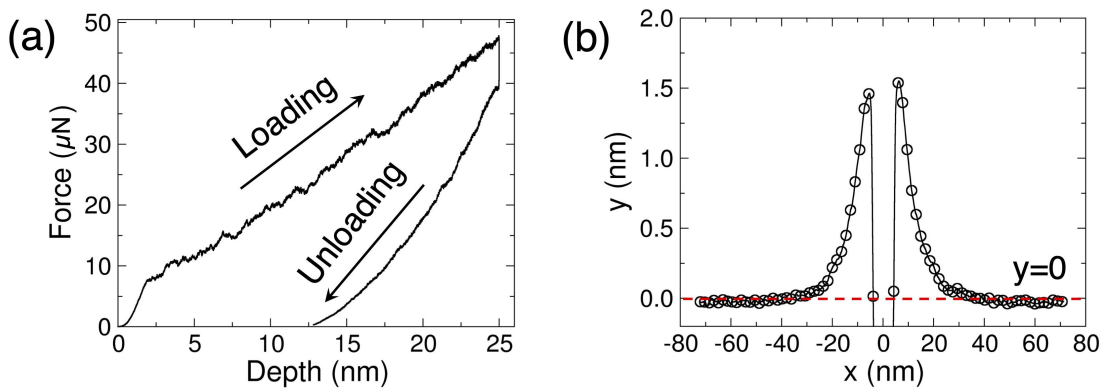

**Figure 2.** (a) Force-displacement curve for the NS3 glass during nanoindentation test. (b) Averaged *y* profile of the NS3 glass after unloading. Error bars (within the symbol size) represent the statistical standard error of the six indentation tests.

## 2.3 Evolution of atomic-level properties of the glass

Microscopic insights into the deformation mechanism of the glass can be obtained by monitoring and quantifying how the various atomic-level structural and mechanical properties evolve during the indentation process. Here, the atomic shear strain introduced by Shimizu *et al.*[47] was used for measuring the local plastic deformation and the atomic stress was measured using the formulation of per-atom stress tensor[48]. Densification is defined as the percentage density change during the deformation with respect to the initial undeformed glass density. To obtain relatively smooth maps of the local properties, we have coarse-grained these atomic-level properties by performing a running average over the spherical radii of 8 Å, 13 Å, and 13 Å for shear strain, shear stress, and densification, respectively.

**Figure 3** shows the maps of shear strain, densification, and the various stress components ($\sigma_{xy}$, $\sigma_{xx}$, $\sigma_{yy}$) at four different stages, i.e., before indentation loading, loading at one-half of the maximum depth, loading at the maximum depth (before relaxation), and after unloading. Firstly, one observes that the shear strain gradually builds up during the loading process and is most pronounced at the upper two corners near the indenter. This concentrated strain quickly decays while moving toward the interior of the glass. After unloading, a large proportion of the shear strain developed during the loading process persists, counting primarily for the irreversible part of the deformation. The evolution of the densification map looks somewhat similar to that of the shear strain, indicating that the changes of these two quantities occur hand in hand[3, 27, 31, 49, 50, 51]. One also gets the impression that, in contrast to the shear strain, most of the densification accumulated during loading seems to have recovered after unloading. However, as will become clear later, this is not really the case as the decay of densification is slower relative to shear strain.

For the shear stress $\sigma_{xy}$, a butterfly-shaped symmetrical pattern is formed along the two sides of the indenter during loading. At the maximum loading depth, compressive stresses are a few GPa in magnitude, and these stresses seem to dissipate at a slower rate when compared with the shear strain and densification. During loading, the glass structures near the indenter tip also experience increasingly strong tensile stress in the *x* direction ($\sigma_{xx}$) and compressive stress in the *y* direction ($\sigma_{yy}$). One observes that the normal stresses can be larger than 10 GPa and even 20 GPa (in the *x* direction) at the maximum loading depth. After unloading, all macroscopic stresses are relaxed away and the atomic stress distribution becomes similar to the initial undeformed state. These observations are in general agreement with previous studies on the nanoindentation of multi-component oxide glasses[13, 24, 27, 32, 52, 53, 54]. Finally, we note that these high stresses in the near-indenter regions are only partially elastic since they induces shear and permanent deformation of the glass structure.

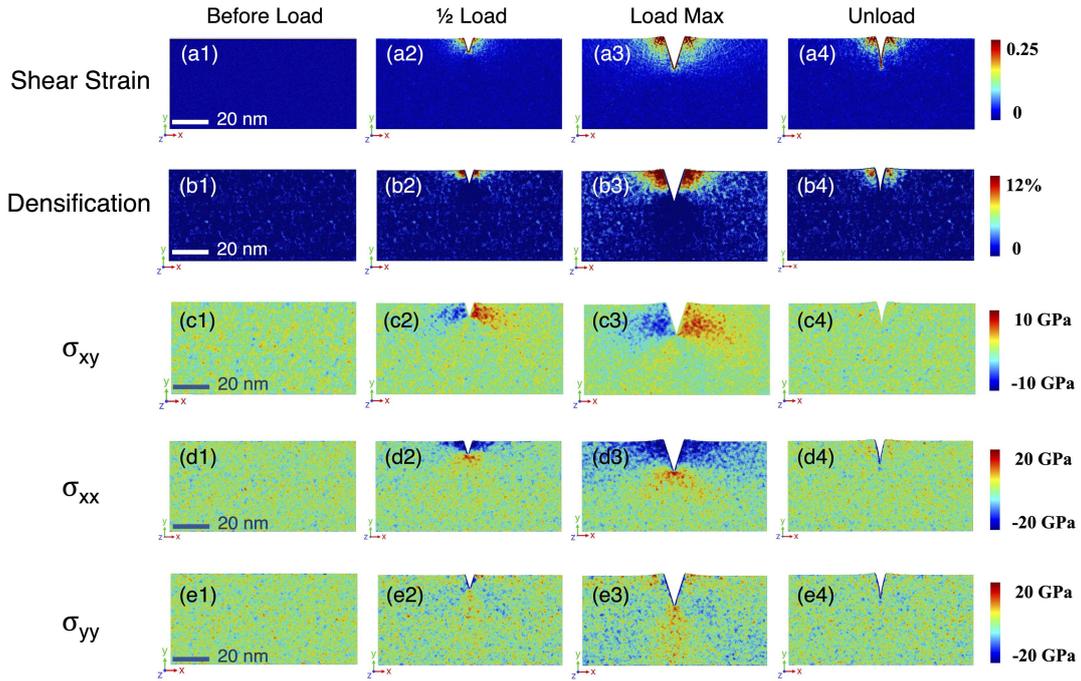

**Figure 3.** Maps of shear strain (a1-a4), densification (b1-b4), shear stress $\sigma_{xy}$ (c1-c4), normal stresses $\sigma_{xx}$ (d1-d4) and $\sigma_{yy}$ (e1-e4) of the NS3 glass at four different deformation stages. From left to right: before loading, loading at one-half of the maximum indentation depth, loading at the maximum depth (before relaxation), and after unloading, respectively.

With a closer inspection of the local maps, it is apparent that the shear strain and densification (i.e., inelastic deformation) extend to a far distance of several tens of nm along the radial directions. In particular, we observe that pronounced shear bands with an interweaving shape exist underneath the indent, penetrating into the sample and covering a large area, see **Fig. 4**. Shear bands can easily progress into shear faults in normal glasses which have been reported in nanoindentation scratch tests of silicate glasses[55], metallic glasses[56, 57] as well as metal-organic framework glasses[58]. The formation of shear bands may be related to the lateral-pushing force as the indenter is pushed into the glass sample.[50, 51, 56, 57, 59, 60, 61]

To reveal the change of atomic structure in a more quantitative manner, we have partitioned off the NS3 glass (after unloading) into small striped-regions (indicated by the dashed lines in **Fig. 4**) parallel to the shear bands (highlighted in the inset of **Fig. 4**). The angle between the striped-regions and the $y$ direction is 60°. (We note that this angle is different from the 45° of shear band as one usually observes in metallic glasses under uniaxial stress since the indentation has made the shear bands reoriented[56].) We have found that varying this partitioning angle by ±15° barely affect on the $d$-dependence of the distribution as well as the decay of these local properties. Hence, the main conclusions reached in this study are expected to be robust.

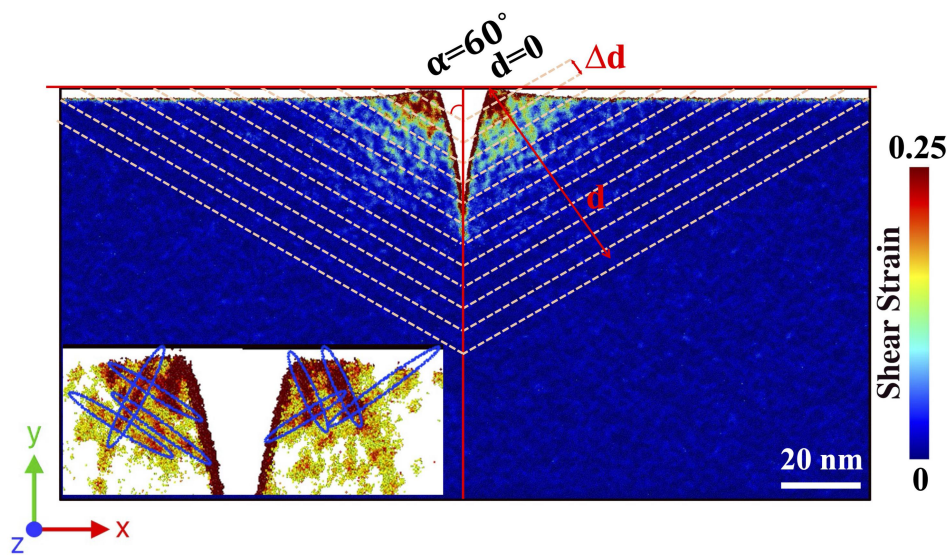

**Figure 4.** Partitioning of the shear strain map after unloading. The orange dash lines represent boundaries that define the striped-regions with a thickness of $\Delta d$. For a given distance $d$ the layer is defined as the region [$d-1/2\Delta d$, $d+1/2\Delta d$]. The inset highlights the interweaving shear bands (denoted by blue ellipses) near the pile-up after unloading.

Having defined the various regions of the glass after unloading, we further probe how the various atomic-level properties evolve with increasing the distance from the immediate contact with the indent. This sort of analysis, which to our knowledge has not been performed in previous studies on the indentation deformation of glasses, is useful for better understanding the evolution of the microscopic properties of the glass due to nanoindentation. In order to eliminate the influence of the free surfaces, we have excluded the atoms in the outermost layer of two nm in thickness.

First, **Fig. 5** presents the probability distribution of the shear strain and densification for the glass regions at different depths with respect to the tip of the pile-up. For shear strain, panel (a), one observes that the distribution profile at $d = 5$ nm is very broad and peaks at ~0.2, indicating that the regions near the indent are severely deformed and exhibit strong heterogeneity. This distribution profile quickly becomes narrower and peaks at smaller values with increasing $d$. In the inset, we show the mean value of the shear strain as a function of *d*. We find that this quantity can be well fitted by an exponential function $y = A \times \exp(-x/\xi)$, where $A$ and $\xi$ are fitting parameters and $\xi$ is also known as the decay length. It is found that the decay length is around 5.3 nm for the shear strain. **Figure 5**b shows the distribution of the densification at various *d*. At *d* = 5 nm, i.e., close to the indent, densification is very pronounced with a peak value of ~8%. With increasing *d*, the degree of densification becomes smaller as the peak value of the distribution profile gradually shifts towards zero. Also in the inset we present the change in the mean value of the densification. Fitting the data using an exponential function shows that the decay length for the densification is around 7.8 nm, which is about 50% larger than that for the shear strain. This result quantitatively indicates that indentation-induced change of glass density actually decays slower than that of shear strain, in contrast to the impression one may gain by visual inspection of the maps of the local properties, see **Fig. 3**.

This finding might be rationalized by considering two effects. First, when under a conjunction of compressive and shear stress, two adjacent part of the glass network can experience irreversible volume shrinkage due to steric hindrance, i.e., the so-called entanglement effect[62]. Second, the thermal activation energy for a densification process may be smaller than the one for shear flow. It was found that for the case of silica glass, the activation energies for shear flow and densification are 511 and 35–55 kJ mol$^{-1}$, respectively.[62, 63] For the NS3 glass studied in this work, we did not find such data in the literature. Nevertheless, it can be expected that the densification process is still easier to be thermally activated than the shear flow, although the difference in activation energy of the two mechanisms might not be as significant as in the case of silica due to the fact that the Na atoms will facilitate local structural rearrangement (and thus shear flow). This means that the change of glass density could extend to larger distances than shear which makes that the decay length for densification is longer than that of shear flow. Thus, while the pile-up (in the immediate vicinity of the indent) has been known to originate from shear flow, our finding suggests that the change of glass density might contribute more than shear flow to the lift-up part as observed in a recent experiment study using a sharp indenter.[13]

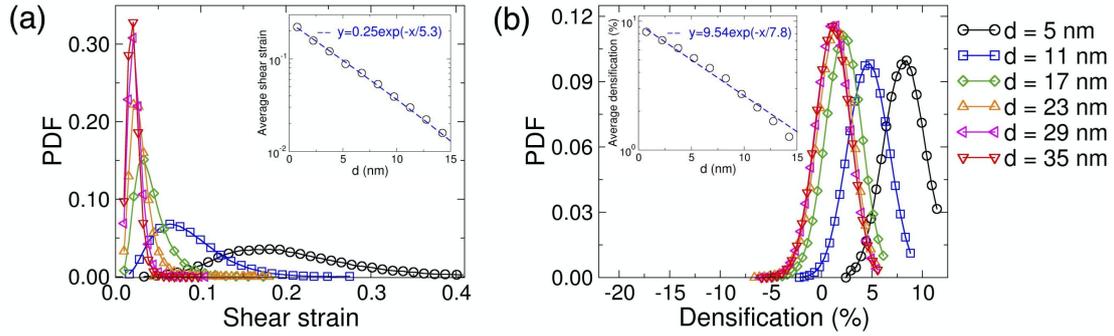

**Figure 5.** Probability density function (PDF) of (a) shear strain and (b) densification at different regions (Δd = 3 nm) of the NS3 glass after unloading. The insets represent the mean values of shear strain and densification versus the distance *d* (log-linear scale). Dashed lines are exponential fits to the data points.

Next, we probe how the atomic structure of the glass evolves with increasing the distance from the indent. We begin by presenting in **Fig. 6** the probability distribution of the most relevant bond lengths and bond angles that are representative of short-range structures. Panel (a) shows that the Si-O bond length is very stable which exhibits very little change with increasing distance *d*. This result could be attributed to the high strength of the Si-O bond. Somewhat surprisingly, the distribution profile of the Na-O bond length also shows a negligible change in the various regions after indentation, as seen in panel (b). This finding might be ascribed to the relatively large mobility (and thus strong adaptability) of the modifiers to find a comfortable bonding environment in the Si-O network.

**Figure 6**c shows that the distribution of the Si-O-Si angle spans a wide range, indicating the large variety of inter-tetrahedral connections. The mean value of the Si-O-Si angle is around 143°, in good agreement with the value of 142° extracted from $^{29}$Si MAS NMR measurements[64]. The dependence of the distribution on *d* is seen in that the main peak is shifted to smaller angles with reducing *d*, indicating that the structure near the indent is more compact due to permanent densification. A further interesting feature is seen at the angle of ~95°, i.e., the occurrence of a small peak corresponding to the presence of two-membered (2M) ring structures (formed by two [SiO$_4$] tetrahedra sharing a common edge). The inset in panel (c) shows the *d*-dependence of the mean Si-O-Si bond angle which can be well-fitted using an exponential function. The corresponding decay length is estimated to be ~9.5 nm, which is slightly larger than the value for the densification,

indicating a close link between the local properties.

For the O-Si-O bond angle, **Fig. 6**d, one observes that the distribution profile is considerably sharper than that of the Si-O-Si angle, indicating that the [SiO$_4$] units are very stable. One nevertheless notices that the main peak at ~110° is lower and slightly broader at small *d*, suggesting that these SiO polyhedra are more distorted near the indenter as a result of severe deformation. Fitting the mean O-Si-O angle (see the inset) using an exponential function shows that the decay length is ~5.9 nm, which is considerably smaller than that of the Si-O-Si angle, indicating that the latter is structurally more flexible in response to deformation.

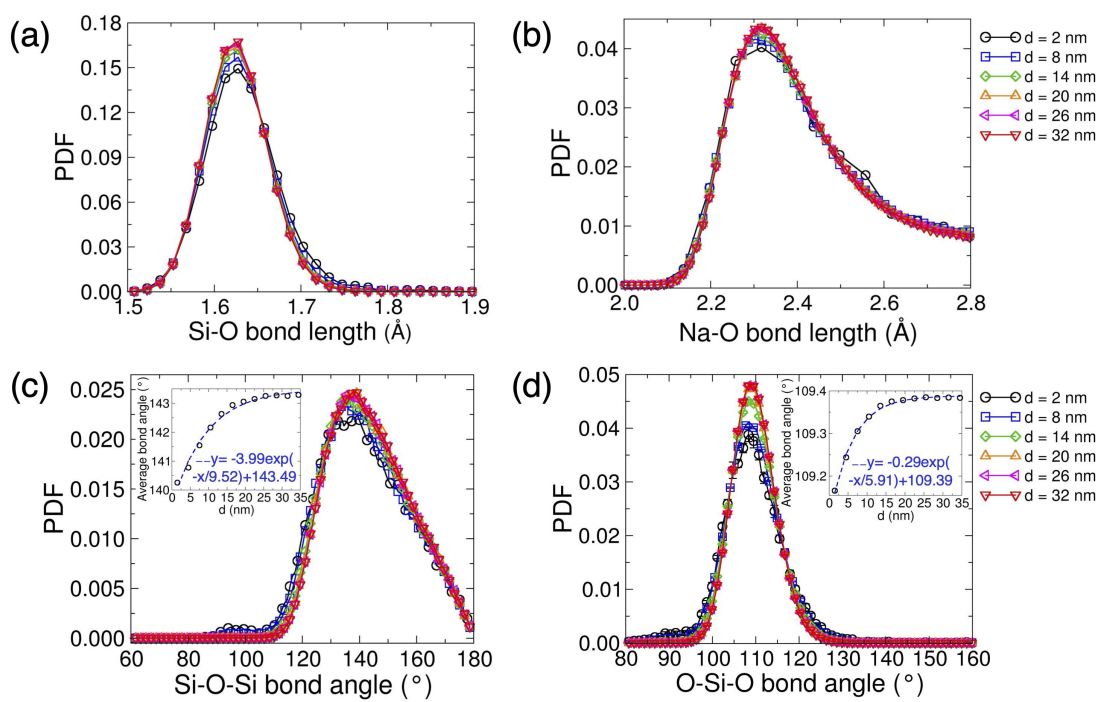

**Figure 6.** Probability density function (PDF) of local structural quantities as functions of (a) Si-O bond length, (b) Na-O bond length, (c) Si-O-Si bond angle, and (d) O-Si-O bond angle at different regions (Δd = 3 nm). The insets in panels (c-d) show the *d*-dependence of the averaged bond angles and the dashed lines are exponential fits to the data.

The change of the short-range structure may propagate to larger distances. To better understand the structural features beyond short range, we probe the change of medium-range structure (MRS) by looking at the primitive ring statistics[65, 66], which was realized using the R.I.N.G.S. code[67]. Here, the size of a ring is denoted as the number of Si nodes the ring contains. **Figure 7** presents the ring size distribution for the regions at different distances from the indent. Firstly, we observe that for all distribution profiles the most probable ring size is five, and its abundance is independent of $d$. Secondly, a pronounced tail towards large ring sizes is seen, indicating that the network is strongly disordered and depolymerized, consistent with the finding of previous studies[33, 68].

Dependence of the ring size distribution on $d$ is seen in both the small- and medium-sized rings. At $d = 2$ nm, i.e., in the immediate vicinity of the indent, one observes that the population of small (2-4 membered) rings is noticeably higher than that in the far-field regions, say $d > 14$ nm. This result can be attributed to the fact that the near-indent regions are severely deformed and densified. Besides, in the near indent regions the population of medium-sized rings (8-11 membered in particular), is higher than that of the far-field regions, indicating that the overall MRS becomes more disordered after indentation. In contrast, the population of the large-sized (> 12-membered) rings seems independent of $d$. Moreover, the wider distribution of the ring size at small $d$ indicates that indentation enhances structural heterogeneity in the glass regions near the indent. As $d$ increases, the degree of deformation rapidly decreases, leading primarily to the reduction of the small- and medium-sized rings. At $d>15$ nm, there is no more discernible change of ring size distribution, indicating that the MRS beyond this distance is not affected by the indentation.

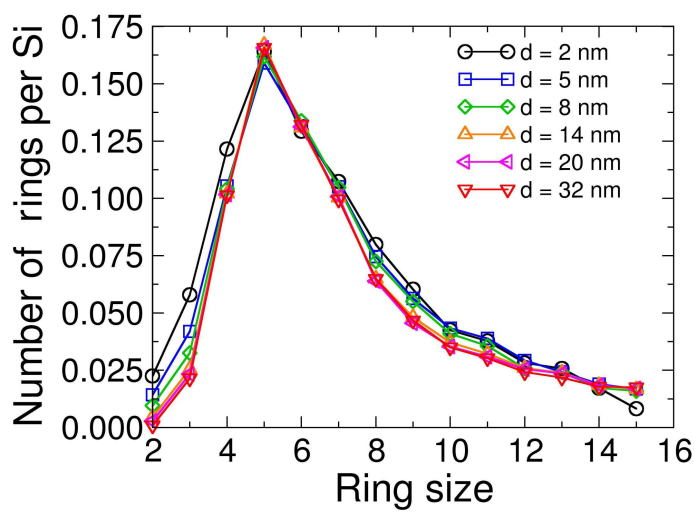

**Figure 7.** Ring size distribution at different regions of the NS3 glass after unloading. Number of rings per Si represents the population of rings normalized by the number of Si. All the results are the averaged values from six parallel tests. Error bars (standard error) are within the symbol size.

Further insight into the indentation-induced variation of glass structure can be obtained by probing how the local coordination environment of different atomic species changes with the distance from the indent, see **Fig. 8**. To allow comparison of the different atomic species, we present the data of coordination number, $Z_{ij}$, i.e., the number of atoms of type $j$ around a central atom of type $i$, as the percentage change with respect to the value of the bulk sample (shown in the bracket). The cutoff distance for determining $Z_{ij}$ corresponds to the first minimum of the partial radial distribution function, $g_{ij}(r)$. (The data presented in Fig. 8 are the averaged coordination number in a given region corresponding to the distance $d$.) For clarity, panel (a) shows the change of $Z$ of the network-forming species, i.e., O and Si, while panel (b) presents the network modifier-related ones. Overall, one recognizes that the changes of $Z$ of the network formers are considerably smaller than those of the network modifiers, indicating that the local coordination environment of Na is much more flexible upon deformation than the local structure of Si and O. This finding is coherent with the fact that Si-O bond is relatively strong and that Na is only weakly bonded to the Si-O network and thus has more freedom to locally rearrange.[35]

Panel (a) shows that the Si-O coordination is very stable; even in the highest deformed state (in the immediate vicinity of the indent) the structural change is within 1% with respect to the bulk glass. Beyond the nearest neighbors, the connection between two tetrahedra, i.e., the Si-Si coordination, is about two times more flexible than the Si-O coordination, consistent with the fact that the change of Si-O-Si angle is the main mechanism for accommodating the deformation.[35, 36] Increasing $d$ leads to the reduction of $Z$ since the structure becomes increasingly less deformed. We find that the changes of $Z$ all decay in an exponential manner and become negligible at $d > 15$ nm.

Panel (b) shows that the change of $Z_{SiNa}$ is most pronounced among all Na-related pairs and that $Z_{NaNa}$ is the least changed one. One notable implication of this result is that SiNa has a stronger tendency to form clusters than NaNa upon indentation deformation. It is recognizable that the changes of Na-related coordination also show an exponentially decaying trend, akin to the other structural measures presented above. To further quantify the decaying behavior of $Z$ we fit the data of SiO and NaO using the exponential function described above. For $Z_{SiO}$ the decay length is around 3.7 nm whereas this value is about 5.0 nm for $Z_{NaO}$, indicating that the change of Na's local environment can propagate to larger distances upon indentation. It is worth noting that the change of Na's coordination environment (which may induce change in network topology) is consistent with the variation of shear strain, implying that the change of the local bonding environment of Na is the structural origin of shear flow in the NS3 glass.

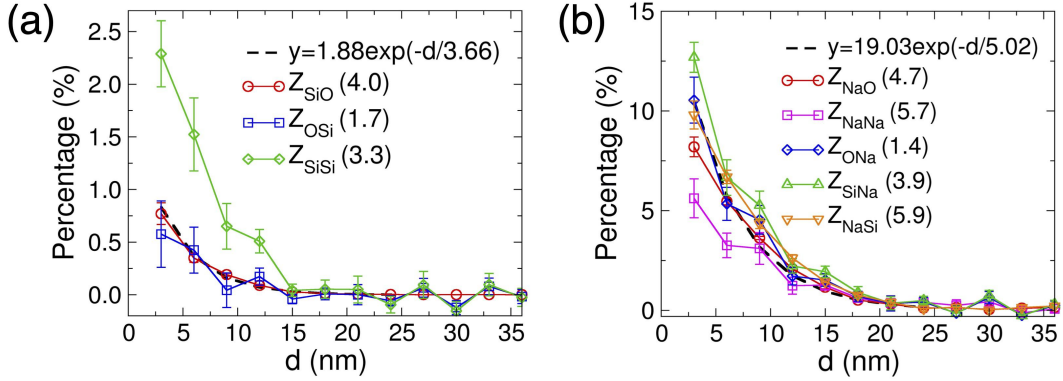

**Figure 8.** (a) Percentage change of the mean coordination number $Z_{ij}$ of the network-forming species as a function of distance $d$ from the indent with respect to the bulk glass. (b) Percentage change of the mean coordination number of the Na-related pairs. All the results are the averaged values from six parallel tests and the error bars represent the statistical standard error of the six parallel indentation tests. The dashed lines are exponential fits to the data for $Z_{SiO}$ and $Z_{NaO}$, respectively. The fitting parameter $y_0$ was set to zero to mitigate the influence from the fluctuation of the data at large $d$.

## 4. Conclusions

Using large-scale MD simulations we have investigated the evolution of atomic-level properties of an archetypal sodium silicate glass under the load of a sharp cube-corner indenter up to an indentation depth of 25 nm. The simulated glass sample is first demonstrated to be realistic as the calculated elastic constants, glass density, and the ratio between the pile-up height and the maximum indentation depth of the simulated glass are in good agreement with experimental measurements.

Shear bands with an interweaving shape are observed underneath the indent. Further quantification of the evolution of the atomic-level properties is achieved through partitioning off the glass samples into localized regions parallel to the shear bands. Moving away from the indent, both shear strain and densification are found to decay in an exponential manner with decay lengths of around 5.3 nm and 7.8 nm, respectively, implying that the latter might contribute more to the lift-up part as recently observed in experiments[13]. The origin of the observed exponential decay might be attributed to the modifiers which enhance local structural flexibility/deformability, making that the perturbation by the surface is damped out quickly. Both the Si-O-Si angle and the population of small-to-medium-sized rings exhibit notable changes below the indent, indicating that indentation deformation induces a more disordered and heterogeneous network structure. The change of coordination number of the network-forming species (i.e., Si and O) is considerably smaller than that of the network modifiers (i.e., Na), indicating that the local coordination environment of Na is much more flexible upon deformation than the local structure of Si and O. The changes of the Si-O-Si angle and Na's coordination environment are responsible for densification and shear flow, respectively, as evidenced from the good correspondence of the decaying behaviors of these local properties.

**Data availability**

All data needed to evaluate the conclusions of this work are present in the paper. Additional data related to this paper may be requested from the authors.

**Declaration of Competing Interest**

The authors declare that they have no known competing financial interests or personal relationships that could have appeared to influence the work reported in this paper.

**Acknowledgments**

We thank Dr. Haidong Liu from Rensselaer Polytechnic Institute for help on the simulation setup and Dr. Clemens Kunisch for valuable comments. The authors gratefully acknowledge financial support from the National Natural Science Foundation of China (52102002, U22A20125), Shanghai Municipal Natural Science Foundation (22ZR1400400), Shanghai Pujiang Program (22PJD002), and the Fundamental Research Funds for the Central Universities (2232022G-07).